\documentclass[baaa]{baaa}

\usepackage[pdftex]{hyperref}
\usepackage{subfigure}
\usepackage{natbib}
\usepackage{helvet,soul}
\usepackage[font=small]{caption}

\contriblanguage{1}

\contribtype{2}

\thematicarea{1}

\title{Cosmology with cosmic voids}

\titlerunning{Cosmology with cosmic voids}

\author{
C.M. Correa\inst{1,2}
\&
D.J. Paz\inst{1,2}
}

\authorrunning{Correa \& Paz}

\contact{cmcorrea@unc.edu.ar}

\institute{
Instituto de Astronom\'ia Te\'orica y Experimental, CONICET--UNC, Argentina
\and
Observatorio Astron\'omico de C\'ordoba, UNC, Argentina
}

\resumen{
Los vacíos cósmicos constituyen laboratorios cosmológicos prometedores.
Sin embargo, resulta fundamental una descripción completa de todos los efectos que sufren las mediciones observacionales en el espacio de redshift con el fin de obtener ajustes cosmológicos no sesgados.
Presentamos una descripción concisa de estos efectos y sentamos las bases teóricas para el diseño de tests cosmológicos confiables basados en la distribución de tamaños de los voids y en la función de correlación cruzada void-galaxia.
Mostramos que los relevamientos espectroscópicos modernos ofrecen una señal ruido alta para detectarlos y estudiarlos.
}

\abstract{
Cosmic voids constitute promising cosmological laboratories.
However, a full description of all the redshift-space effects that affect observational measurements is mandatory in order to obtain unbiased cosmological constraints.
We make a description in a nutshell of these effects and lay the theoretical foundations for designing reliable cosmological tests based on the void size function and the void-galaxy cross-correlation function.
We show that modern spectroscopic surveys offer a high signal-to-noise ratio to detect and study them.
}


\begin{document}

\maketitle


\section{Introduction}
\label{sec:intro}

Cosmic voids are promising cosmological probes for testing the dark energy problem and alternative gravity theories.
Since they are vast underdense regions of space, they encode key information about the geometry and expansion history of the Universe.
Moreover, the potential of voids has increased recently with the development of modern spectroscopic surveys, which are covering a volume and redshift range without precedents.

There are two primary statistics in void studies: (i) the void size function, that describes the abundance of voids \citep{svdw,abundance_jennings}, and (ii) the void-galaxy cross-correlation function \citep{clues2,rsd_cai}, that characterises the density and peculiar velocity fields around them.
Both statistics are affected by distortions in the observed spatial distribution of galaxies, which translate into anisotropic patterns: (i) the redshift-space distortions \citep[RSD]{rsd_kaiser}, a dynamical effect, and (ii) the \citet[AP]{ap} distortions, a geometrical effect.
Both effects encode fundamental information about the cosmological parameters \citep{aprsd_correa, aprsd_hamaus2020, rsd_hawken2, aprsd_nadathur2020}.

However, there has been an obstacle that has prevented the use of voids as reliable cosmological probes.
Our standard picture of distortions is incomplete.
Traditionally, we have focused only on the spatial distribution of galaxies around voids.
The truth is that the RSD and AP effects also affect intrinsic void properties, such as their number, their size and their spatial distribution.
These systematics generate additional anisotropic patterns on observations that lead to biased cosmological constraints if they are not taken into account properly.

The purpose of this article is to compile the main results of various works in order to lay the theoretical foundations for designing reliable and unbiased cosmological tests based on the void size function and the void-galaxy cross-correlation function.


\section{Redshift-space effects in voids}
\label{sec:zeffects}

This section is based on the results obtained in \cite{zvoids_correa} using the so-called spherical void finder \citep{clues3}.
Galaxies can be considered as particles under the mapping between real and redshift space.
They are totally conserved in this process, only their position changes.
Voids, by contrast, are extensive regions: some may get lost, others may be created artificially.
Nevertheless, we found that voids above the shot-noise level are almost conserved under this mapping, the loss of voids decreases as larger sizes are considered.
Therefore, it is statistically valid to assume \textit{void number conservation}.

Redshift-space voids are systematically bigger than their real-space counterparts.
This \textit{expansion effect} (hereafter t-RSD) is a consequence of the RSD induced by tracer dynamics at scales around the void radius.
The volume of voids is also affected by the AP effect.
This is because the discrepancies between the fiducial and true cosmologies generate deviations between the void dimensions along ($\parallel$) and perpendicular ($\perp$) to the line of sight (hereafter LOS).
This is the \textit{Alcock-Paczynski volume effect}.
An analytical description is possible from dynamical (Eq.~\ref{eq:velocity}) and cosmological (Eq.~\ref{eq:ap_coord}) considerations, from which a linear relation is found between the real and redshift-space radii: $R_{\rm v}^{\rm rs}$ and $R_{\rm v}^{\rm zs}$, respectively:
\begin{equation}
    R_{\rm v}^{\rm zs} = q_{\rm AP} ~ q_{\rm RSD} ~ R_{\rm v}^{\rm rs},
    \label{eq:q_ap_rsd}
\end{equation}
where
\begin{equation}
    q_{\rm RSD} = 1 - \frac{1}{6} \frac{f(z)}{b} \Delta_{\rm v, g}^{\rm NL}, \\
    q_{\rm AP} = \sqrt[3]{(q_{\rm AP}^\perp)^2 q_{\rm AP}^\parallel},
    \label{eq:q_factors}
\end{equation}
where in turn, $q_{\rm AP}^\perp = D_{\rm M}^{\rm fid}(z)/D_{\rm M}^{\rm true}(z)$ and $q_{\rm AP}^\parallel = H_{\rm true}(z)/H_{\rm fid}(z)$.
In $q_{\rm RSD}$, $\Delta_{\rm v, g}^{\rm NL}$ is the average density contrast threshold defined for void identification, $f(z)$ the logarithmic growth rate of density perturbations, and $b$ the bias parameter relating the galaxy and matter density fields.
In $q_{\rm AP}$, $D_{\rm M}$ denotes the comoving angular diameter distance, and $H$ the Hubble parameter.
The indices ``fid" and ``true" refer to fiducial and true cosmology quantities, respectively.
In both cases, $z$ is the mean redshift of the void.

Finally, redshift-space void centres are systematically shifted along the LOS.
This is the \textit{off-centring effect} (hereafter v-RSD), a consequence of a new class of RSD induced by the global dynamics of the region containing the void.
An analytical description can be obtained from an RSD displacement applied to the void centre:
\begin{equation}
    s_{\rm v \perp} = r_{\rm v \perp}, \\
    s_{\rm v \parallel} = r_{\rm v \parallel} + v_{\rm v \parallel} \frac{1 + z}{H(z)}.
	\label{eq:vrsd}
\end{equation}
Here, $\mathbf{r}_{\rm v} = (r_{\rm v \perp}, r_{\rm v \parallel})$, $\mathbf{s}_{\rm v} = (s_{\rm v \perp}, s_{\rm v \parallel})$ and $\mathbf{v}_{\rm v} = (v_{\rm v \perp}, v_{\rm v \parallel})$ denote the real-space position vector of the centre, its redshift-space counterpart, and the void net velocity \citep{lambas_sparkling_2016}, respectively.


\section{Void size function}
\label{sec:vsf}

The void size function quantifies the comoving number density of cosmic voids as a function of their effective radii.
It is analogous to the mass function of dark matter haloes.
Therefore, it can be modelled using the excursion set formalism combined with the spherical evolution of density perturbations \citep{svdw}.

Following this approach, \cite{abundance_jennings} derived a \textit{volume conserving model}, which assumes that the total volume occupied by cosmic voids is conserved in the transition from linear to non-linear regime:
\begin{equation}
    \frac{dn_{\rm v}}{d\mathrm{ln}R_{\rm v}^{\rm rs}} = 
    \frac{f_{\mathrm{ln}\sigma}(\sigma)}{4/3 \pi (R_{\rm v}^{\rm rs})^3} \frac{d\mathrm{ln}\sigma^{-1}}{d\mathrm{ln}R_{\rm v, L}^{\rm rs}}
    \frac{d \mathrm{ln} R_{\rm v, L}^{\rm rs}}{d \mathrm{ln} R_{\rm v}^{\rm rs}}.
    \label{eq:vsf_vdn}
\end{equation}
Here, $\sigma$ is the square root of the mass variance (dependent on the matter power spectrum), whereas $f_{\mathrm{ln}\sigma}$ is the fraction of the Universe occupied by voids:
\begin{equation}
    f_{\mathrm{ln}\sigma} = 2 \sum_{j=1}^{\infty} j\pi \chi^2 \mathrm{sin}(j\pi \mathcal{D}) \mathrm{exp} \left[ - \frac{(j\pi \chi)^2}{2} \right],
    \label{eq:flnsig}
\end{equation}
where in turn, $\chi = \mathcal{D}\sigma / |\Delta_{\rm v}^{\rm L}|$ and $\mathcal{D} = |\Delta_{\rm v}^{\rm L}| / (|\Delta_{\rm v}^{\rm L}| + \Delta_{\rm c}^{\rm L})$.
$\Delta_{\rm c}^{\rm L}$ and $\Delta_{\rm v}^{\rm L}$ represent the two barriers needed in the excursion set to take into account both the void-in-cloud and void-in-void modes.
The former is expected to vary within $1.06 < \Delta_{\rm c}^{\rm L} < 1.686$, the moments of turn-around and collapse, respectively.
The latter is expected to be the moment of shell-crossing in the expansion process.
Finally, $R_{\rm v, L}^{\rm rs}$ is the linear radius predicted by theory.
It can be related to its non-linear counterpart by a constant factor: $R_{\rm v}^{\rm rs} = \gamma R_{\rm v, L}^{\rm rs}$, with $\gamma = (1 + \Delta_{\rm v}^{\rm NL})^{-1/3}$.

The shell-crossing moment for voids identified from the underlying total matter distribution corresponds to the non-linear value $\Delta_{\rm v}^{\rm NL} = -0.8$, or equivalently, to the linearly extrapolated value $\Delta_{\rm v}^{\rm L} = -2.71$.
However, in practice, we can only define voids from the observed galaxy distribution.
In this case, it is not trivial to relate this event to a specific density threshold.
This can be overcome by the procedure suggested by \cite{abundance_bias_contarini}.
First, we define an observational density threshold as low as possible: $\Delta_{\rm v, g}^{\rm NL}$ (usually less than $-0.7$).
Then, we relate this value to the one corresponding to the total matter density field by means of the bias parameter: $\Delta_{\rm v}^{\rm NL} = \Delta_{\rm v, g}^{\rm NL}/b$.
Finally, we derive its corresponding linear value with the fitting formula provided by \cite{bernardeau}: $\Delta_{\rm v}^{\rm L} = C[1-(1+\Delta_{\rm v}^{\rm NL})^{-1/C}]$, with $C=1.594$.

The last aspect that we need to consider is to relate the redshift and real-space void radii by means of the t-RSD and AP volume effects (Eq.~\ref{eq:q_ap_rsd}).
These are the theoretical foundations for designing a cosmological test with the void size function.


\section{Void-galaxy cross-correlation function}
\label{sec:vgcf}

\begin{figure}[!t]
\centering
\includegraphics[width=85mm]{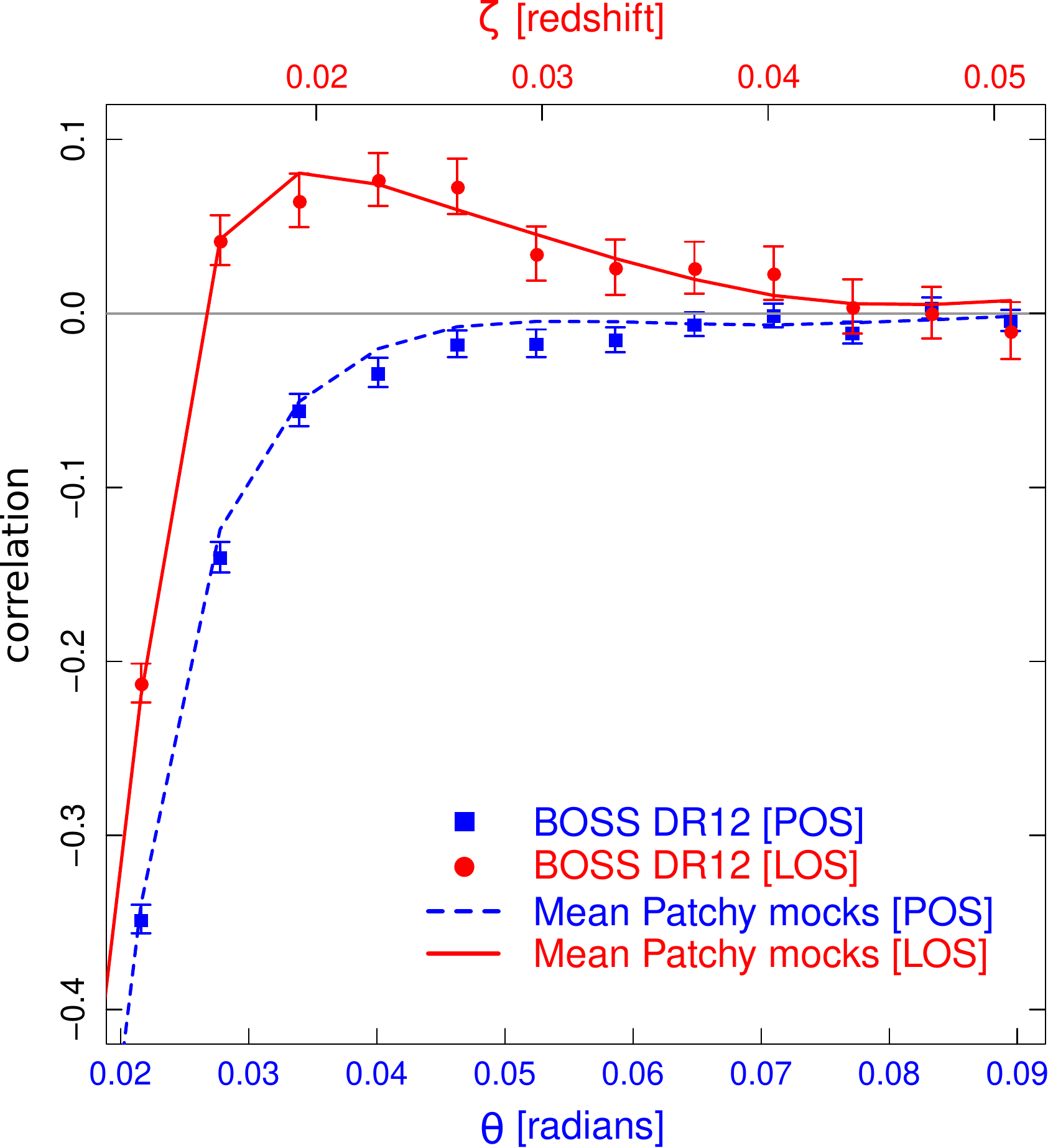}
\caption{
Plane-of-sky and line-of-sight projections of the void-galaxy cross-correlation function measured from BOSS DR12 data and from the simulated MultiDark Patchy mocks.
}
\label{fig:boss}
\end{figure}

In \cite{aprsd_correa}, we provided an empirical fitting expression for the real-space correlation function suitable for voids above the shot-noise level:
\begin{equation}
  \xi(r) =
    \begin{cases}
      Ar - 1 & \text{if} ~ r < r_\mathrm{wall}\\
      -\xi_0 \left[ \left( \frac{r}{r_0} \right)^{-3} + \left( \frac{r}{r_0} \right)^{-\alpha} \right] & \text{if} ~ r \geq r_\mathrm{wall},
    \end{cases}       
  \label{eq:dens_model}    
\end{equation}
where $r$ denotes distance to the void centre, and $r_{\rm wall}$ is a characteristic scale separating their inner and outer parts.
This model has three free parameters: $\xi_0$, $r_0$ and $\alpha$ (the slope A can be derived from $\Delta_{\rm v, g}^{\rm NL}$).

To study the distortion effects, it is more convenient to represent the correlation function as a 2D contour map with axes perpendicular and along the LOS: $\xi(s_\perp,s_\parallel)$.
The following expressions relate the real and redshift-space separation vectors between a void-galaxy pair: $\mathbf{r} = (r_\perp, r_\parallel)$ and $\mathbf{s} = (s_\perp, s_\parallel)$, via their relative velocity: $\mathbf{v} = (v_\perp, v_\parallel)$:
\begin{equation}
    s_\perp = r_\perp, \\
    s_\parallel = r_\parallel + v_\parallel \frac{1+z}{H(z)}.
	\label{eq:rsd_coord}
\end{equation}

The t-RSD effect can be modelled with the so-called \textit{Gaussian streaming model} \citep{clues2}:
\begin{equation}
    \resizebox{0.9\hsize}{!}{
    $1 + \xi(s_\perp, s_\parallel) = 
    \int_{-\infty}^{\infty} \frac{1 + \xi(r)}{\sqrt{2 \pi}\sigma_{\rm v}}
    \mathrm{exp} \left[- \frac{(v_\parallel - v(r)\frac{r_\parallel}{r})^2}{2\sigma_{\rm v}^2} \right]
    \mathrm{d}v_\parallel$,
    }
    \label{eq:gsm}
\end{equation}
where $v(r)$ is a profile characterising the velocity field $\mathbf{v}$, and $\sigma_{\rm v}$ the velocity dispersion, taken as a free parameter.
The former can be derived analytically in the linear regime:
\begin{equation}
    v(r) = - \frac{H(z)}{(1+z)} \frac{f(z)}{b} \frac{1}{r^2} \int_0^r \xi(r') r'^2 dr'.
	\label{eq:velocity}
\end{equation}

Actually, the distance between a void-galaxy pair must be inferred from an angle $\theta$ on the plane of the sky (hereafter POS) and a redshift separation $\zeta$ along the LOS.
This involves the following cosmological equations:
\begin{equation}
    s_\perp = D_{\rm M}(z) \theta, \\
    s_\parallel = \frac{c}{H(z)} \zeta,
	\label{eq:ap_coord}
\end{equation}
where $c$ denotes the speed of light.
We consider in this way the AP effect.

Now, we will follow the methodology that we developed in \cite{aprsd_correa} by considering the POS and LOS projections of the correlation function.
They are obtained respectively by projecting it towards the $\perp$-axis in a given redshift range, and towards the $\parallel$-axis in a given angular range.
It is important to consider that correlations are measured via a binning scheme, where several scales are mixed.
A generic bin is a cylindrical shell oriented along the LOS with the following dimensions: $s_\perp^{\rm int}$ (internal radius), $s_\perp^{\rm ext}$ (external radius), $s_\parallel^{\rm low}$ (lower height) and $s_\parallel^{\rm up}$ (upper height).
Then, the correlation estimator for such a bin is given by:
\begin{equation}
    \xi_{\rm bin} = 2 
    \frac{ \int_{s_\parallel^{\rm low}}^{s_\parallel^{\rm up}} ds_\parallel \int_{s_\perp^{\rm int}}^{s_\perp^{\rm ext}} s_\perp [1 + \xi(s_\perp,s_\parallel)] d s_\perp }
    { ((s_\perp^{\rm ext})^2-(s_\perp^{\rm int})^2) (s_\parallel^{\rm up}-s_\parallel^{\rm low}) } - 1.
    \label{eq:scales}
\end{equation}
The projected correlations are special cases with the following limits: (i) $s_\parallel^{\rm low} \rightarrow 0$ and $s_\parallel^{\rm up} \rightarrow \mathrm{PR_\parallel}$ for the POS correlation, and (ii) $s_\perp^{\rm int} \rightarrow 0$ and $s_\perp^{\rm ext} \rightarrow \mathrm{PR_\perp}$ for the LOS correlation.
Here, $\mathrm{PR_\perp}$ and $\mathrm{PR_\parallel}$ are the corresponding projection ranges.

This model still needs to consider the v-RSD effect.
This is work in progress.
It can be modelled by incorporating information about the void net velocity distribution into Eq.~(\ref{eq:gsm}).
But there is more.
In \cite{zvoids2_correa}, we found that there is an extra source of anisotropic patterns in the correlation function: the intrinsic ellipsoidal nature of voids.
This is the \textit{ellipticity effect} (e-RSD).
Currently, we are working in an extension of the spherical void finder that allows to define voids as free-shape underdense regions of maximal volume (Paz et al, in prep.).
These are the theoretical foundations for designing a cosmological test with the correlation function.

By way of illustration\footnote{More details can be found at Dr. Correa's PhD thesis \url{https://rdu.unc.edu.ar/handle/11086/21041}.}, Fig.~\ref{fig:boss} shows the POS and LOS projections of the correlation function measured from the spectroscopic survey BOSS DR12 \citep{boss}.
We also show the results obtained from the simulated MultiDark Patchy mock surveys \citep{patchy_kitaura}.
The consistency between both data sets is a promising result.
The error bars were calculated from the covariance matrix obtained from these mocks.
In both cases, we selected a void sample with sizes between $30 \leq R_\mathrm{v}^{\rm zs}/\mathrm{Mpc} \leq 35$.
Correlations were treated directly in terms of the angular and redshift coordinates, $\theta$ and $\zeta$, hence it is not necessary to assume a fiducial cosmology for the posterior analysis.
We took the following projection ranges: $\mathrm{PR}_\perp = 0.0232$ and $\mathrm{PR}_\parallel = 0.0131$, chosen to be approximately equal to $30~\mathrm{Mpc}$ in the fiducial cosmology.
This allows us to capture effectively the anisotropic patterns in both directions.
The LOS projection is much more sensitive to the distortion effects than the POS projection.
Hence, the large differences between the two confirm the fact that it is possible to detect and analyse all the redshift-space effects addressed in this work with high precision.


\section{Conclusions}
\label{sec:conclusions}

Cosmic voids are promising cosmological probes.
However, it is mandatory a full description of all the redshift-space effects that affect observational measurements in order to obtain unbiased cosmological constraints.
Void systematics can be summarised as follows: (i) the AP effect, (ii) the expansion effect (t-RSD), (iii) the off-centring effect (v-RSD), and (iv) the ellipticity effect (e-RSD).
We made a description in a nutshell of these effects and laid the theoretical foundations for designing reliable cosmological tests with the void size function and the void-galaxy cross-correlation function.


\bibliographystyle{baaa}
\small

\end{document}